\documentclass[pra,floatfix,showpacs,amsmat,twocolumn]{revtex4}
\usepackage{amssymb}
\usepackage{graphicx}
\usepackage{amsmath}
\usepackage{colordvi}
\usepackage{bbm}
\usepackage{theorem}
\newtheorem{definition}{Definition}

\newtheorem{lemma}{Lemma}
\newtheorem{theorem}{Theorem}
\newtheorem{corollary}{Corollary}

\begin{document}

\title{Negative entanglement measure for bipartite separable mixed states}
\author{Cheng-Jie Zhang$^{1,2}$}
\author{Yong-Jian Han$^1$}
\author{Yong-Sheng Zhang$^1$}
\email{yshzhang@ustc.edu.cn}
\author{Yu-Chun Wu$^1$}
\email{wuyuchun@ustc.edu.cn}
\author{Xiang-Fa Zhou$^1$}
\author{Guang-Can Guo$^1$}
\affiliation{$^1$Key Laboratory of Quantum Information, University of
Science and Technology of China, Hefei, Anhui 230026, People's
Republic of China\\
$^2$Department of Physics, National University of Singapore, 2 Science Drive 3, 117542, Singapore}

\begin{abstract}
We define a negative entanglement measure for separable states which shows that how much entanglement one should compensate the unentangled state at least for changing it into an entangled state. For two-qubit systems and some special classes of states in higher-dimensional systems, the explicit formula and the lower bounds for the negative entanglement measure have been presented, and it always vanishes for bipartite separable pure states. The negative entanglement measure can be used as a useful quantity to describe the entanglement dynamics and  the quantum phase transition. In the transverse Ising model, the first derivatives of negative entanglement measure diverge on approaching the critical value of the quantum phase transition, although these two-site reduced density matrices have no entanglement at all. In the 1D Bose-Hubbard model, the NEM as a function of $t/U$ changes from zero to negative on approaching the critical point of quantum phase transition.
\end{abstract}


\maketitle

\section{Introduction}
After a fundamental nonclassical aspect of entanglement was recognized by Einstein, Podolsky, and Rosen in 1935 \cite{Bell}, Werner proposed the accurate definitions of separability and entanglement in 1989 \cite{werner}. In recent years, with the rapid progress in quantum information science, it has become more and more clear that entanglement is a valuable quantum resource \cite{nielsen} and acts as an important role in many other physical phenomenon such as quantum phase transition \cite{QPT,QPT2} and efficient simulation of many body systems \cite{vidal}. Therefore, many entanglement measures have been proposed in order to quantify it \cite{review,formation,relative,robust,negativity,squashed}. However, separability and entanglement are not yet fully understood \cite{review1,review2,escon,Oleg,Ma,Li,Kotowski}, and it is also necessary to present measures for separable states to show how unentangled the state is.

Why to define separability measure? One reason is that in many practical situations, people need to know how can we transform given separable states to entangled states. However, for typical entanglement measures, separable states always have zero entanglement. To distinguish the difference of each separable state, one needs a separability measure to show how unentangled a given separable state is. Another reason is for the \textit{integrity} of the entanglement measure theory. The separability measure is a natural complementarity for entanglement measure and can present detailed descriptions of separable states which, in the following, will be analytically explained in the two-qubit system. Thirdly, quantitative investigation of entanglement dynamics involves both entangled and separable states. This is evident when we consider the sudden death and birth of entanglement in open quantum systems \cite{Yu}. Similar situation occurs in the investigation of quantum phase transition by entanglement measures.

In this work, we define a negative entanglement measure (NEM), i.e., separability measure for separable states which shows that how much entanglement one should pay at least for changing the unentangled state into an entangled state. Based on this definition, edge states of separable (ESS) states are also introduced to describe the boundary between the separable and entangled states, and the necessary and sufficient condition of two-qubit ESS states can be obtained via the NEM. Furthermore, for two-qubit systems, we obtain an explicit formula for the NEM, and exact values or lower bounds for special classes of states in higher-dimensional systems are also presented, and the NEM always vanishes for bipartite separable pure states. Interestingly, the NEM can be used to describe the quantum phase transition of the transverse Ising model and the 1D Bose-Hubbard model with suitable parameters. The first derivative of NEM in the transverse Ising model diverges on approaching the critical value for the next-next-nearest neighboring sites, although this two-site reduced density matrix has no entanglement. And the NEM as a function of $t/U$ in the 1D Bose-Hubbard model changes from zero to negative on approaching the critical point of quantum phase transition.

Another application is for entanglement dynamics \cite{Yu}. Actually, a pioneer research using negative concurrence for entanglement dynamics has been done by Yu and Eberly \cite{Eberly}.

\section{Definitions}
Consider a bipartite system where the dimension of its Hilbert space $\mathcal{H}$ is finite, we can define the NEM for separable states.

\begin{definition}
For a bipartite separable state $\rho\in\mathcal{H}$, its negative entanglement is
\begin{equation}\label{negative}
    N(\rho)=-\inf_{\{t,\sigma\}}\{tC(\sigma):C(\frac{\rho+t\sigma}{1+t})>0\},
\end{equation}
where the infimum is taken over all possible $t$ and $\sigma$ ($t>0$, $\sigma\in\mathcal{H}$ is an arbitrarily entangled bipartite state). $C$ denotes the entanglement measure concurrence or I concurrence for two-qubit system or higher-dimensional bipartite system, respectively.
\end{definition}

It is noticed that every separable state has non-positive value of NEM. The absolute value of NEM $|N(\rho)|$ describes how much entanglement at least a separable state $\rho$ should mix with, in order to wipe out all separability of $\rho$. On the contrary, any entangled state (not necessarily normalized) with less than $|N(\rho)|$ entanglement must be changed into a separable state by mixing with $\rho$.

The definition of the NEM can be viewed as a parallel definition of robustness of entanglement \cite{robust}, since $|N(\rho)|$ shows the robustness of separability. Similar to robustness of entanglement, $|N(\rho)|$ describes the robustness of the separability of $\rho$, i.e. the amount of entanglement at least the separable state $\rho$ should mix with in order to wipe out all separability. Therefore, the NEM has its physical meaning and it can be regarded as complement to the entanglement measures, e.g. for describing quantum phase transition and entanglement dynamics.

The same idea can be used to define the ESS states.
\begin{definition}
A separable state $\rho\in\mathcal{H}$ is an edge state of separable states if, for arbitrary $\epsilon>0$, there exists an entangled state $\sigma\in\mathcal{H}$ such that $(\rho+\epsilon\sigma)/(1+\epsilon)$ is entangled.
\end{definition}

\begin{figure}
\begin{center}
\includegraphics[scale=0.5]{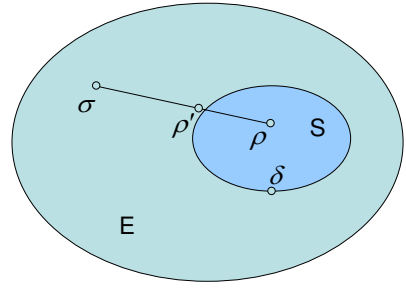}
\caption{(Color online) The set of separable states S is represented by the inner ellipse, and the outer ellipse stands for the set of all states. Entangled states E are out of the inner ellipse and in the outer ellipse. A separable state $\rho$ is changed into an entangled state $\rho'$ by mixing with an entangled state $\sigma$, and the ESS state $\delta$ exactly lies in the boundary between S and E.}\label{1}
\end{center}
\end{figure}

Fig. \ref{1} demonstrates the NEM and the ESS states. A separable state $\rho$, mixing with an entangled state $\sigma$, is changed into another entangled state $\rho'=(\rho+t\sigma)/(1+t)$. The infimum amount of mixed entanglement is defined as $|N(\rho)|$. It is also shown in Fig. \ref{1} that the ESS state $\delta$ exactly lies in the boundary between the separable and the entangled states. Furthermore, according to Definitions 1 and 2, one can obtain the following properties.

(i) $N(\rho)=0$ for every ESS state $\rho$.

(ii) $N(\rho)=N(U_{L}\rho U_{L}^{\dag})$ for any local unitary operator $U_{L}=U_{1}\otimes U_{2}$.

(iii) $N(\rho)\leq\sum_{k}p_{k}N(\rho_{k})$, where $\rho=\sum_{k}p_{k}\rho_{k}$, $\{\rho_{k}\}$ are separable states and $\sum_{k}p_{k}=1$.

\section{Two-qubit system}
The concurrence for a two-qubit state $\rho$ is $C(\rho)=max\{\Lambda(\rho),0\}$, where $\Lambda(\rho)=\lambda_{1}-\lambda_{2}-\lambda_{3}-\lambda_{4}$, and the
$\lambda_{i}$s are squared roots of eigenvalues of $\rho(\sigma_{y}\otimes\sigma_{y})\rho^{\ast}(\sigma_{y}\otimes\sigma_{y})$ in the decreasing order \cite{concurrence}. According to Refs. \cite{frank1,frank2,frank3}, $\Lambda(\rho)$ can also be written as $\Lambda(\rho)=-1/2 \min_{L_{1},L_{2}}\mathrm{Tr}(L_{1}RL_{2}^{T})$, where $[R]_{ij}=\mathrm{Tr}(\rho\sigma_{i}\otimes\sigma_{j})$ with $\{\sigma_{j}\}$ the Pauli spin matrices, and $L_{1}$, $L_{2}$ are proper orthochronous Lorentz transformations. Using this variational characterization, one can obtain that $\Lambda(\rho)$ is a convex function \cite{frank1,frank2,frank3}.

\begin{lemma}
$\sum_{i}p_{i}\Lambda(\rho_{i})\geq\Lambda(\sum_{i}p_{i}\rho_{i})$, where $0\leq p_{i}\leq1$ and $\sum_{i}p_{i}=1$.
\end{lemma}
Proof. Suppose that $\mathrm{Tr}[L_{1}(\sum_{i}p_{i}R_{i})L_{2}^{T}]$ reaches its minimum when $L_{1}=\mathcal{L}_{1}$ and $L_{2}=\mathcal{L}_{2}$. Therefore, $\min_{L_{1},L_{2}}\mathrm{Tr}[L_{1}(\sum_{i}p_{i}R_{i})L_{2}^{T}]=\sum_{i}p_{i}\mathrm{Tr}(\mathcal{L}_{1}R_{i}\mathcal{L}_{2}^{T})\geq\sum_{i}p_{i}[\min_{L_{1}^{i},L_{2}^{i}}\mathrm{Tr}(L_{1}^{i}R_{i}{L_{2}^{i}}^{T})]$.
Since  $\Lambda(\rho)=-1/2 \min_{L_{1},L_{2}}\mathrm{Tr}(L_{1}RL_{2}^{T})$ and  $\Lambda(\rho_{i})=-1/2 \min_{L_{1}^{i},L_{2}^{i}}\mathrm{Tr}(L_{1}^{i}R_{i}{L_{2}^{i}}^{T})$, one can obtain that $\sum_{i}p_{i}\Lambda(\rho_{i})\geq\Lambda(\sum_{i}p_{i}\rho_{i})$ holds.  \hfill $\square$

Interestingly, for two-qubit systems, the NEM defined in Eq. (\ref{negative}) has a close relation with $\Lambda(\rho)$.
\begin{theorem}
For a two-qubit separable state $\rho$, the exact value of NEM is
\begin{equation}\label{}
    N(\rho)=\Lambda(\rho).
\end{equation}
\end{theorem}
Proof. Since $\sigma$ is an arbitrarily entangled two-qubit state, $C((\rho+t\sigma)/(1+t))>0$ and $C(\rho)=max\{\Lambda(\rho),0\}$, one can obtain that $C((\rho+t\sigma)/(1+t))=\Lambda((\rho+t\sigma)/(1+t))$ and $C(\sigma)=\Lambda(\sigma)$. According to Lemma 1, the convex inequality is $\Lambda(\rho)/(1+t)+t\Lambda(\sigma)/(1+t)\geq\Lambda((\rho+t\sigma)/(1+t))>0$. Therefore, $tC(\sigma)=t\Lambda(\sigma)>-\Lambda(\rho)$ holds. In the following, it will be proved that the infimum of $tC(\sigma)$ is indeed $-\Lambda(\rho)$.

In Ref. \cite{concurrence}, every two-qubit separable state $\rho$ has the decomposition $\rho=\sum_{i=1}^{4}|x_{i}\rangle\langle x_{i}|$ where $\langle x_{i}|\widetilde{x}_{j}\rangle=\lambda_{i}\delta_{ij}$ and $|\widetilde{x}_{j}\rangle\equiv\sigma_{y}\otimes\sigma_{y}|x_{j}^{\ast}\rangle$. (i) If $\lambda_{1}>0$, for arbitrary $\epsilon>0$, one can choose $\sigma=|x_{1}\rangle\langle x_{1}|$ and $t=(\lambda_{2}+\lambda_{3}+\lambda_{4}-\lambda_{1}+\epsilon)/\lambda_{1}$ ($t>0$ for separable states), such that $tC(\sigma)=\lambda_{2}+\lambda_{3}+\lambda_{4}-\lambda_{1}+\epsilon=-\Lambda(\rho)+\epsilon$ and $C((\rho+t\sigma)/(1+t))=\epsilon/(1+t)>0$ hold. (ii) If $\lambda_{1}=0$, $\lambda_{2}=\lambda_{3}=\lambda_{4}=0$ and $-\Lambda(\rho)=0$ as well. For arbitrary $\epsilon>0$, one can choose $\sigma=|\psi^{+}\rangle\langle\psi^{+}|$ ($|\psi^{+}\rangle=(|00\rangle+|11\rangle)/\sqrt{2}$) and $t=\epsilon>0$, such that $tC(\sigma)=\epsilon=-\Lambda(\rho)+\epsilon$ and $C((\rho+t\sigma)/(1+t))=t/(1+t)>0$ hold. Therefore, the infimum of $tC(\sigma)$ is indeed $-\Lambda(\rho)$ and the explicit formula for the NEM in two-qubit systems is $N(\rho)=\Lambda(\rho)$. \hfill $\square$

Remark. Theorem 1 states that the quantity of entanglement a two-qubit separable state $\rho$ should mix with is at least $|\Lambda(\rho)|$, in order to wash out all the separability. In other words, the quantity of separability of $\rho$ is $\Lambda(\rho)$. Moreover, Theorem 1 also points out the relation between ESS states and $N(\rho)=0$ from its proof.

\begin{corollary}
In two-qubit systems, a separable state $\rho$ is an ESS state if and only if
\begin{equation}\label{necessary}
  N(\rho)=0.
\end{equation}
\end{corollary}
Proof. If a separable state $\rho$ satisfies $N(\rho)=0$, from the proof of Theorem 1 there are two situations. (i) If $\lambda_{1}>0$, for arbitrary $\epsilon>0$, $t=\epsilon/\lambda_{1}>0$, there is an entangled state $\sigma=|x_{1}\rangle\langle x_{1}|$ such that $C((\rho+t\sigma)/(1+t))>0$. (ii) If $\lambda_{1}=0$, for arbitrary $\epsilon>0$, $t=\epsilon>0$, one can choose $\sigma=|\psi^{+}\rangle\langle\psi^{+}|$ such that $C((\rho+t\sigma)/(1+t))>0$. Therefore, if a separable state $\rho$ satisfies $N(\rho)=0$, it must be an ESS state. The converse of the Corollary is immediate from Definitions 1 and 2.  \hfill $\square$

Remark. For higher-dimensional systems, an ESS state $\rho$ must satisfy $N(\rho)=0$, i.e., Eq. (\ref{necessary}) is still a necessary condition of ESS states.

\begin{corollary}
For two-qubit separable state $\rho$, $|N(\rho)|$ does not increase on average under local operations and classical communication (LOCC) operations.
\end{corollary}
Proof. This corollary can be proved directly following the proof of Theorem 3 in \cite{frank1}.  \hfill $\square$

Remark. Corollary 2 shows a similar property as an entanglement monotone.

Let us demonstrate Theorem 1 with a simple example. Consider the separable state $\rho=a|00\rangle\langle00|+b|01\rangle\langle01|+c|10\rangle\langle10|+d|11\rangle\langle11|$,
where $0\leq a,b,c,d\leq1$ and $a+b+c+d=1$. According to Theorem 1, its NEM is
\begin{eqnarray}
N(\rho)=\left\{
\begin{array}{ll}
-2\sqrt{bc}, &\quad    ad\geq bc,\\
-2\sqrt{ad}, &\quad    ad<bc.
\end{array}\right.\label{}
\end{eqnarray}
One can choose a special class of entangled states as $\sigma$, i.e., $\sigma(\theta)=\cos\theta|00\rangle+\sin\theta|11\rangle$. When $t>t_{0}(\theta)=\sqrt{8bc/(1-\cos4\theta)}$, the final state $\rho'=(\rho+t\sigma)/(1+t)$ is entangled. Therefore, in the special class of $\sigma$, $\min_{\theta}t_{0}(\theta)C(\sigma(\theta))=2\sqrt{bc}$. (i) If $ad\geq bc$, $\min_{\theta}t_{0}(\theta)C(\sigma(\theta))=|N(\rho)|$, i.e., $t_{0}(\theta_{min})$ and $\sigma(\theta_{min})$ are just the minimum of all $t>0$ and entangled states $\sigma$. (ii) If $ad<bc$, $\min_{\theta}t_{0}(\theta)C(\sigma(\theta))>|N(\rho)|$. $t_{0}(\theta_{min})$ and $\sigma(\theta_{min})$ cannot reach the minimum. Actually, one can choose $\sigma=(\sqrt{b}|01\rangle+\sqrt{c}|10\rangle)/\sqrt{b+c}$. When $t>t_{0}=(b+c)\sqrt{ad/bc}$, $\rho'$ is entangled, and $t_{0}C(\sigma)=2\sqrt{ad}=|N(\rho)|$. It means that $t_{0}$ and $\sigma$ reach the minimum. Numerical simulation also shows that the infimum of $tC(\sigma)$ is $|N(\rho)|$ for all $t>0$ and entangled states $\sigma$ to get an entangled final state $\rho'$.

Remark. According to Definition 1, arbitrarily close to some separable states are entangled states, and others are not. That is because we consider all the states in the Hilbert space $\mathcal{H}$ of the given separable state $\rho$, and the same consideration has also been used in robustness of entanglement. For example, given a two-qubit separable state $\rho$, we calculate Eq. (1) using all the entangled two-qubit states and find the infimum among all entangled two-qubit states. Therefore, we obtain the infimum among all entangled states in the Hilbert space of $\rho$, which is similar to robustness of entanglement.

\section{Higher-dimensional bipartite system}
The \textit{I} concurrence of a bipartite pure state is defined as $C(|\psi\rangle)\equiv\sqrt{2(1-\mathrm{Tr}\rho_{A}^{2})}$,  where the reduced density matrix $\rho_{A}$ is obtained by tracing over the subsystem B \cite{concurrence1}. Furthermore, the definition of \textit{I}
concurrence can be extended to mixed states $\rho$ by the convex
roof, $C(\rho)=\inf_{\{p_{i},|\psi_{i}\rangle\}}
\sum_{i}p_{i}C(|\psi_{i}\rangle),
    \   \rho=\sum_{i}p_{i}|\psi_{i}\rangle\langle\psi_{i}|$,
for all possible decomposition into pure states, where $p_{i}\geq0$
and $\sum_{i}p_{i}=1$. Although the generally explicit formula for \textit{I} concurrence is still under research, for pure states and some special classes of states such as Isotropic states the explicit formulas are available. In the following, we also present exact values and lower bounds of NEM for pure separable states and separable Isotropic states, respectively.

\begin{theorem}
The NEM for pure separable states in bipartite systems is
\begin{equation}\label{}
    N(\rho)=0.
\end{equation}
\end{theorem}
Proof. For arbitrary pure separable state $\rho=|\phi_{1}\rangle\langle\phi_{1}|\otimes|\phi_{2}\rangle\langle\phi_{2}|$, one can select a new basis where $|\phi_{1}\rangle\equiv|0_{1}\rangle$ and $|\phi_{2}\rangle\equiv|0_{2}\rangle$. The pure separable state $\rho$ can be written as $\rho=|0\rangle\langle0|\otimes|0\rangle\langle0|$ and one can choose $\sigma=|\psi^{+}\rangle\langle\psi^{+}|$ under the new basis. Using the partial transposition criterion \cite{Peres}, it can be obtained that the final state $(\rho+t\sigma)/(1+t)$ is entangled for arbitrary $t>0$, i.e., $C((\rho+t\sigma)/(1+t))>0$. Therefore, the infimum of $tC(\sigma)$ when $\sigma=|\psi^{+}\rangle\langle\psi^{+}|$ is $0$. On the other hand, $N(\rho)\leq0$ for all separable states. Hence, $N(\rho)=0$ holds for arbitrary pure bipartite separable state $\rho$. \hfill $\square$

Remark. Theorem 2 shows that pure separable states can be changed into entangled states by mixing with an arbitrarily tiny amount of entanglement. From the proof, one can see that every separable pure state in bipartite systems is an ESS state.

Isotropic states are a class of mixed states for $d\times d$ systems, which can be expressed as $\rho_{F}=(I-|\Psi^{+}\rangle\langle\Psi^{+}|)(1-F)/(d^{2}-1)+F|\Psi^{+}\rangle\langle\Psi^{+}|$, where $|\Psi^{+}\rangle=1/\sqrt{d}\sum_{i=1}^{d}|ii\rangle$, $F=\langle\Psi^{+}|\rho_{F}|\Psi^{+}\rangle$ satisfying $0\leq F\leq1$. Rungta and Caves \cite{rungta} have given an explicit formula for \textit{I} concurrence of isotropic states,
\begin{eqnarray}
C(\rho_{F})=\left\{
\begin{array}{ll}
0, & 0\leq F\leq1/d,\\
\sqrt{2d/(d-1)}(F-1/d), & 1/d\leq F\leq1.
\end{array}\right.\label{Crho}
\end{eqnarray}
According to Definition 1, one can choose $\sigma=|\Psi^{+}\rangle\langle\Psi^{+}|$, and get a lower bound of NEM of separable isotropic states,
\begin{equation}\label{lower}
    N(\rho_{F})\geq \sqrt{2d/(d-1)}(F-1/d),
\end{equation}
when $0\leq F\leq1/d$. Interestingly, the formula of this lower bound Eq. (\ref{lower}) is just the same as concurrence of entangled isotropic states in Eq. (\ref{Crho}).

\begin{figure}
\begin{center}
\includegraphics[scale=0.7]{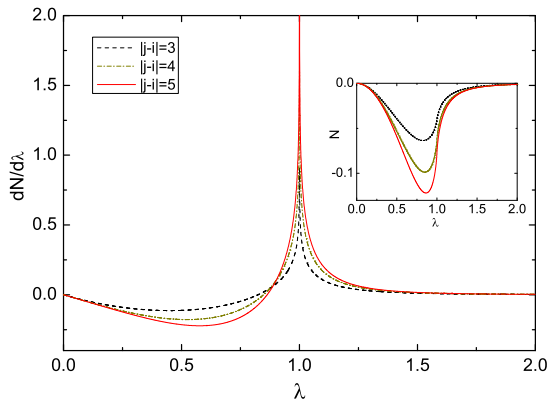}
\caption{(Color online) Quantum phase transition of the Ising model is described by $dN/d\lambda$ when $|j-i|=3,4,5$. Although these two sites have no entanglement, their first derivatives of NEM as a function of $\lambda$ still diverge on the critical point $\lambda_{c}=1$.}\label{2}
\end{center}
\end{figure}

\section{Applications}
It has been shown that entanglement can be used as a useful quantity to describe quantum phase transition
and entanglement dynamics. However, in many specific physical models, the ground states are not always
entangled, which greatly limits the scope of entanglement measures in these systems. This shortcoming can
be amended by NEM. Actually, NEM can be used in the investigation of quantum phase transition and entanglement
dynamics even when there is no entanglement at all.

In Refs. \cite{QPT,QPT2}, the concurrence of nearest-neighboring sites and next-nearest-neighboring sites has been used to exhibit a quantum phase transition in the XY model. The Hamiltonian of $N$ sites on a 1D lattice with cyclic boundary conditions is
\begin{equation}\label{XY}
    H=-\sum_{j=0}^{N-1}\bigg(\frac{\lambda}{2}[(1+\gamma)\sigma_{j}^{x}\sigma_{j+1}^{x}+(1-\gamma)\sigma_{j}^{y}\sigma_{j+1}^{y}]+\sigma_{j}^{z}\bigg),
\end{equation}
where $\sigma_{j}^{a}$ is the Pauli matrix ($a=x,y,z$) at site $j$. When $\gamma=1$, Eq. (\ref{XY}) reduces to the Ising model. In Refs. \cite{QPT,QPT2}, the concurrence of the reduced density matrix $\rho(i,j)$  has been considered, where $i,j$ are positions of two spins, and $\rho(i,j)=\rho(|j-i|)$ because of translation invariance. It is shown that only nearest-neighboring sites ($|j-i|=1$) and next-nearest-neighboring sites ($|j-i|=2$) have entanglement, and all other pairs have zero two-party entanglement in the Ising model \cite{QPT,QPT2}. However, we can use the NEM to describe all other pairs. In the thermodynamic limit, i.e. $N\rightarrow\infty$, the reduced density matrix $\rho(i,j)$ can be obtained by one- and two-point correlation functions, which have been shown in \cite{correlation1,correlation2,correlation3}. Therefore, one can easily calculate the NEM according to Theorem 1. Fig. \ref{2} shows the NEM and its first derivative as a function of $\lambda$ of two sites when $|j-i|=3,4,5$. These two sites have no entanglement indeed, and the first derivatives $dN/d\lambda$ diverge on approaching the critical value $\lambda_{c}=1$, which are similar with $dC(1)/d\lambda$ shown in \cite{QPT}. Therefore, the NEM can be regarded as complement to the concurrence.

\begin{figure}
\begin{center}
\includegraphics[scale=0.8]{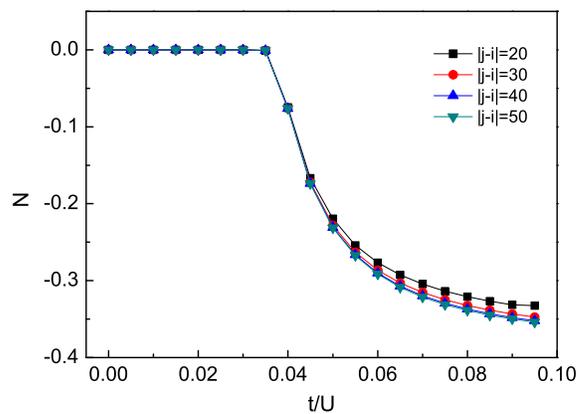}
\caption{(Color online) Quantum phase transition of the Bose-Hubbard model is described by the NEM when $|j-i|=20,30,40,50$. Although these two sites have no entanglement, their NEM as a function of $t/U$ changes from zero to negative on approaching the critical point $t/U=0.035$.}\label{3}
\end{center}
\end{figure}

Another example of investigating quantum phase transition by the NEM is the 1D Bose-Hubbard model with suitable parameters. The Hamiltonian for a 1D lattice of $N$ sites has the form
\begin{equation}\label{}
    H=-t\sum_{\langle i,j\rangle}(b_i^{\dag}b_j+b_j^{\dag}b_i)+\frac{U}{2}\sum_{i=1}^{N}\hat{n}_i(\hat{n}_i-1)-\mu \sum_{i=1}^{N}\hat{n}_i,
\end{equation}
where $b_i^{\dag}$ creates a particle in site $i$ and $\hat{n}_i=b_i^{\dag}b_i$, $t$ is the hopping energy of bosons between neighbouring lattice sites $\langle i,j\rangle$, $U$ is the on-site repulsion between particles, and $\mu$ is the chemical potential. In the thermodynamic limit $N\rightarrow\infty$ and translation symmetry situation, the reduced density matrix $\rho(i,j)$ can be numerically simulated by the iTEBD algorithm, a recently developed method related to density matrix renormalization group techniques \cite{tebd1,tebd2}. If the parameter $\mu/U$ is small, the NEM can be used to investigate quantum phase transition in the 1D Bose-Hubbard model.
Fig. \ref{3} shows the NEM as a function of $t/U$ when $\mu/U=0.08$ and $|j-i|=20,30,40,50$. These two sites have no entanglement indeed, and the NEM changes from zero to non-zero on approaching the critical value $t/U=0.035$. Therefore, the NEM can be regarded as complement to the concurrence.

The NEM can also be used in entanglement dynamics \cite{Yu}. Yu and Eberly have done a pioneer research of two-qubit entanglement dynamics using negative concurrence \cite{Eberly}, which they used is actually NEM according to Theorem 1. They have shown that NEM can signal whether the disentanglement of an entangled state occurs in a finite time, and the final value of NEM generally depends on the system's initial state under fully disentangling evolution, whereas a disentangled concurrence must have the value $0$ no matter what the initial state was. Therefore, one can obtain much more information in the evolution using NEM and concurrence than using concurrence only.

\section{Discussion and Conclusion}
The discussions in this work can be interestingly generalized. From Definition 1, it is realized that one can define NEM for multipartite separable states by choosing a proper multipartite entanglement measure instead of concurrence in Eq. (\ref{negative}). Furthermore, for bipartite separable states, we can also use other entanglement measures instead of $C$ in Definition 1, such as the entanglement of formation \cite{formation}, relative entropy of entanglement \cite{relative}. The different entanglement measures describe different ways to quantify entanglement, and $|N(\rho)|$ always denotes, under the chosen way of entanglement quantification, how much entanglement at least the separable state $\rho$ should mix with to wipe out all separability of $\rho$. Interestingly, one can use a special entanglement measure $E(\rho)=0$ or $1$ for separable or entangled states, respectively, and Definition 1 becomes a parallel definition of robustness of entanglement \cite{robust}. Last but not least, it is worth noticing that the relative entropy is a well defined distance measure \cite{relative}, and it would give an alternative way of defining a negative entanglement measure, by using the distance to the closest entangled state. Unfortunately, it is quite difficult to obtain the exact value of negative entanglement measure using relative entropy, even in the two-qubit system. Therefore, using relative entropy, a similar theorem to Theorem 1 is not available at present, but it is interesting and worth for further research.

In conclusion, we have defined the NEM for separable states which shows that how much entanglement one should pay at least for changing the unentangled state to an entangled state. The quantification describes how unentangled every separable state is. Another definition of ESS states has been given to describe the boundary between the separable and entangled states, and we have also shown the close relation between ESS states and the NEM. Furthermore, one can obtain an explicit formula for the NEM in two-qubit systems, and exact values or lower bounds for special class of states in higher-dimensional systems, and the NEM always vanishes for bipartite separable pure states. Interestingly, the proposed measures can also be used to investigate entanglement dynamics and quantum phase transition in many-body systems, although in this case the considered system has no entanglement. Entanglement is not the only quantity which signals the critical point of a quantum phase transition. Therefore, detailed discussion of the separability and entanglement in this specific physical systems will give a deeper insight into such novel phenomena.

\section*{ACKNOWLEDGMENTS}
We thank Prof. Shunlong Luo and Prof. Zhengwei Zhou for helpful discussions and the anonymous referee for suggestions. This work was funded by the National Fundamental Research Program (Grant No. 2006CB921900), the National Natural Science Foundation of China (Grant Nos. 10674127, 60621064, 10974192 and 10974193), the Innovation Funds from the Chinese Academy of Sciences, and the K.C. Wong Foundation. C.J.Z. acknowledges the financial support of ASTAR Grant R-144-000-189-305.

\end{document}